\documentclass[preprintnumbers,amsmath,superscriptaddress]{revtex4}
\usepackage{graphicx}
\usepackage{amsmath}
\usepackage{amssymb}
\usepackage{epsfig}
\usepackage{amsthm}
\usepackage{color}
\def\be{\begin{equation}}
\def\ee{\end{equation}}
\def\arr{\begin{array}{rll}}
\def\ea{\end{array}}
\def\bea{\begin{eqnarray}}
\def\eea{\end{eqnarray}}
\begin{document}
\title{Maxwell fish eye  for polarized light}
\author{Mher Davtyan}
\affiliation{Institute of Radiophysics and Electronics, Ashtarak-2, 0203, Armenia }
\email{mher.davtyan@gmail.com}
\author{Zhyrair Gevorkian}
\affiliation{Institute of Radiophysics and Electronics, Ashtarak-2, 0203, Armenia }
\affiliation{Yerevan Physics Institute, 2 Alikhanian Brothers St.,  0036, Yerevan, Armenia}
\email{gevork@yerphi.am}
\author{Armen Nersessian}
\affiliation{Yerevan Physics Institute, 2 Alikhanian Brothers St., 0036, Yerevan, Armenia}
\affiliation{Yerevan State University, 1 Alex Manoogian St., Yerevan, 0025, Armenia}
\affiliation{Bogoliubov Laboratory of Theoretical Physics, Joint Institute for Nuclear Research, Dubna, Russia}
\email{arnerses@yerphi.am}
\begin{abstract}
We consider the propagation of polarized light in the medium with Maxwell fish eye refraction index profile. We show that polarization violates the additional symmetries of the medium so that ray trajectories no longer remain closed. Then we suggest a modified, polarization-dependent  Maxwell fish eye refraction index which restores all symmetries of the initial profile and yields closed trajectories of polarized light. Explicit expressions for the polarization-dependent integrals of motion and the solutions of corresponding ray trajectories are presented.
\end{abstract}
\maketitle

\section{Introduction}
It is well-known that the minimal action principle came in physics from geometric optics. Initially, it was invented for the description of the propagation of light and is presently known as the Fermat principle
\begin{equation}
 {\cal S}_{Fermat} =\frac{1}{\lambdabar_{0}}\int d{\tilde l},\qquad d{\tilde l}:=n(\mathbf{ r}) |d\mathbf{r}/d\tau |d\tau
\label{gactions2}
\end{equation}
where  $n(\mathbf{ r})$ is the
refraction index, and $\lambdabar_0$ is the wavelength in vacuum. This  action
could be interpreted as the action of the system on the
three-dimensional curved space equipped with the ``optical metrics" or the Fermat metrics (see. \cite{arnold})
of Euclidean signature
 \begin{equation}
  d{\tilde l}^2= n^2(\mathbf{r})d\mathbf{r} \cdot d\mathbf{r}\;.
\label{om}
\end{equation}
Thus, the symmetries of the system which describe the propagation of light in a particular medium are coming from the symmetries of the optical metrics of that particular medium. In systems with a maximal number of functionally independent integrals of motion ($2N-1$ integrals for $N$-dimensional system), all the trajectories of the system become closed.
The closeness of the trajectories makes such profiles highly relevant in the study of cloaking and perfect imaging phenomena.
The most well-known profile of this sort is the so-called ``Maxwell fish eye" profile which is defined by the metrics of (three-dimensional) sphere or pseudosphere (under pseudosphere we mean the upper (or lower) sheet of the two-sheet hyperboloid).
\begin{equation}
n_{Mfe}(\mathbf{r}) =\frac{n_0}{|1 +  \kappa  \mathbf{r}^2 |}, \qquad \kappa=\pm\frac{1}{4r^2_0}
\label{Mfe0}
\end{equation}
where the plus/minus sign in the expression for $\kappa$ corresponds to the sphere/pseudosphere with the radius  $r_0$, and $n_0>0$.
Apart from applications in cloaking and perfect imaging phenomena \cite{Pendry06,Leonhardt06,leonhardconformal}, Maxwell fish eye is a common profile in quantum optics with single atoms and photons \cite{perczel2018}, optical resonators \cite{turks14}, discrete spectrum radiation \cite{discretespectrumradiation} etc.
Moreover,  there are many experimental implementations of the  Maxwell fish eye lenses \cite{zhao2021bifunctional,lu2020millimeter,koala2020}.

In most of the listed studies, scalar wave approximation was used and polarization of electromagnetic waves was not taken into account. In these systems, the ray trajectories belong to the plane which is orthogonal to the angular momentum.  Introduction of spin (polarization) results to the rotation of this plane by a constant angle proportional to spin, moreover,  it breaks the non-rotational symmetries of the optical systems with Maxwell fish eye profile, so that photon trajectories no longer remain closed   \cite{gevdav20}. Thus, the key property of the Maxwell fish-eye profile which makes it relevant in cloaking and perfect imaging phenomena is violated.

In the present paper, we continue our study of a polarized light passing through the Maxwell fish eye profile within the geometrical optics approximation. The key point of our study is that we  introduce a   polarization-dependent  deformation of the Maxwell fish-eye profile
\be
n^{s}_{Mfe}(\mathbf{r})= \frac{n_{Mfe}(\mathbf{r})}{2} \left(1 + \sqrt{1 - \frac{4\kappa s^2\lambdabar^2_0 }{n_0}\frac{1}{n_{Mfe}(\mathbf{r})}}\right),
\label{spinFishEye0}
\ee
where $n_{Mfe}(\mathbf{r})$ is the original Maxwell fish eye profile given by \eqref{Mfe0}, and $s$ is the light polarization. For the linearly and circularly  polarized light we have $s=0$ and $s=1$ respectively.
The proposed deformation restores all the symmetries of the optical Hamiltonian, with Maxwell fish eye profile, which were broken after the inclusion of polarization. It also ensures the closeness of the trajectories for polarized photons and can be used for cloaking and perfect imaging of polarized photons.
It is seen, that spin induced term is proportional to the dimensionless parameter $s^2\lambda_0^2/r_0^2$ where $r_0$ is the characteristic length of the profile defined in \eqref{Mfe0}. This means that spin will play a significant role only in the vicinity of wave and geometrical optics border $s\lambda_0/r_0\sim 1$, since  we are working in the framework of geometrical optics approximation, $\lambda_0\ll r_0$. Below we will investigate the influence of spin (polarization) on the ray trajectories  in the deformed  Maxwell fish eye profile given by \eqref{spinFishEye0}.\\

The paper is organized as follows. In {\sl Section 2}, we describe the  Hamiltonian formulation of the geometric optical system given by the action \eqref{gactions2}.
We also present some other textbook fact on the duality between Coulomb and free-particle system on a (pseudo)sphere which allows to relate the Maxwell fish eye and Coulomb profiles and will be used in our further consideration.

In {\sl Section 3}, we present the Hamiltonian formalism for the polarized light propagating in an optical medium and propose the general scheme of the deformation of isotropic refraction index profile which allows us to restore the initial symmetries after the inclusion of polarization.

In {\sl Section 4}, we use the proposed scheme for the construction of "polarized Maxwell fish eye"  profile \eqref{spinFishEye0} which inherits all the symmetries of the original profile \eqref{Mfe0} when light polarization is taken into account. We present the explicit expressions for the symmetry generators of the corresponding Hamiltonian system and find the expressions of the Casimirs of their symmetry algebra.

In {\sl Section 5} the explicit expressions for the trajectories of polarized light are presented. It is shown that these trajectories are no longer orthogonal to the angular momentum but turn to a fixed angle relative to it.  Despite deviations from circles, these trajectories remain closed.\\
%
%

Through the text we will use the notation $r:=|\mathbf{r}| $, $\mathbf{r}:=(x_1,x_2,x_3)$, $\mathbf{p}:=(p_1,p_2,p_3)$, $p:= |\mathbf{p}| $, and so on.

\section{Scalar waves}

Due to reparametrization-invariance of the action   \eqref{gactions2},  the Hamiltonian constructed by the standard Legendre transformation is identically zero. However, the constraint between  momenta  and coordinates appears there
\begin{equation}
\Phi:= \frac{{\bf p}^2}{n^2({\bf r})} -\lambdabar^{-2}_0 =0.
\label{constraint}
\end{equation}
Hence, in accordance with the Dirac's constraint theory \cite{dirac}
the respective  Hamiltonian system
is defined by the
canonical Poisson brackets
\begin{equation}
\{x_i, p_j\}=\delta_{ij},\quad   \{p_i, p_j\}=\{x_i, x_j\} =0,
\label{pb0}\end{equation}
and by the  Hamiltonian
\begin{equation}
\mathcal{H}_0=\alpha({\bf p},{\bf r})\Phi =\alpha({\bf p},{\bf r})\left( \frac{{  p}^2}{n^2({\bf r})}-\lambdabar^{-2}_0 \right)\approx 0.
\label{h0}
\end{equation}
Here $\alpha$ is the Lagrangian  multiplier  which could be an arbitrary function of coordinates and momenta, and $i,j=1,2,3$.
The notation ``weak zero", $\mathcal{H}_{0}\approx 0$,  means that when writing down the Hamiltonian equations of motion,
we should take into account the constraint \eqref{constraint} only after the differentiation,
\be
\frac{df({\bf r},{\bf p})}{d\tau} =\{f, \mathcal{H}_0\}=\{f,\alpha\}\Phi+\alpha \{f,\Phi\}\approx\alpha\{f,\Phi\}.\ee
The  arbitrariness in the choice of the function $\alpha$ reflects the reparametrization-invariance of \eqref{gactions2}.
For the description of the  equations of motion in terms of arc-length of the original Euclidian space one should choose (see,  e.g.  \cite{bliokh})
\begin{equation}
\alpha =\frac{n^2({\bf r})}{ {  p} + \lambdabar^{-1}_0 n({\bf r})},\qquad \Rightarrow\quad \mathcal{H}_{\rm Opt}= p - \lambdabar^{-1}_0n({\bf r}).
\label{alpha}
\end{equation}
With this choice, the  equations of motion take the conventional form \cite{ko}
\begin{equation}
\frac{d{\bf {p}}}{dl}=\lambdabar^{-1}_0{\bf \nabla} n({\bf r}),\qquad
\frac{d{\bf {r}}}{dl}=\frac{\bf p}{ {  p} },
\label{hameq}
\end{equation}
where $dl:=\alpha({\bf r}, {\bf p})d\tau$ is the element of arc-length.
These equations describe the motion of a wave package with center coordinate ${\bf r}$ and momentum ${\bf p}$ in the medium with refraction index $n({\bf r})$.\\

Assume we have a Hamiltonian system given by the Poisson bracket \eqref{pb0} and by the Hamiltonian
\be
H=\frac{{  p}^2}{2g({\bf r})}+V({\bf r}).
\label{h}\ee
In accordance with the Mopertuit principle, after fixing the energy surface $H=E$, we can relate its trajectories with the optical Hamiltonian \eqref{h0} with the refraction index
\be
 n( {\bf r})=\lambdabar_0\sqrt{2g({\bf r})(E-V({\bf r}))}.
\label{rih} \ee
Clearly, the optical Hamiltonian \eqref{h0} (as well as the Hamiltonian  \eqref{alpha}) with the  refraction index  \eqref{rih} inherits all the symmetries and constants of motion of the Hamiltonian \eqref{h}.

Canonical transformations preserve the symmetries of the Hamiltonians and their level surfaces. Hence, we are able to construct the physically non-equivalent optical Hamiltonians (and refraction indices) with the identical symmetry algebra.
The simplest illustration is the well-known relation between the Coulomb Hamiltonian which defines the so-called Coulomb refraction index profile
and the free-particle Hamiltonian on the three-dimensional sphere, which defines the ``Maxwell fish eye" refraction index (see e.g. \cite{perelomov}).
Firstly, we fix the energy surface of the  Coulomb Hamiltonian and get the respective refraction index
\begin{equation}
H_{Coul}-E :=\frac{{ p}^2}{2}-\frac{{\gamma }}{  r }-E=0, \quad \Rightarrow \quad n_{Coul}=\lambdabar_0\sqrt{2(E+\gamma/r ) },\qquad {\rm where}\quad \gamma>0.
\label{Coulsur}\end{equation}

The constants of motion of the Coulomb problem (and of the respective optical Hamiltonian) are given by the rotational momentum and by the Runge-Lenz vector
\be
\mathbf{L}=\mathbf{r}\times \mathbf{p},\qquad {\bf A} = {\bf L }\times\mathbf{p} + \gamma  \frac{{\bf r}}{  {r} }
\label{A}\ee
which form the algebra
\be
 \{A_i,A_j\}=-2\varepsilon_{ijk}H_{Coul}L_k,\quad \{A_i,L_j\}=\varepsilon_{ijk}A_k,\quad \{L_i,L_j\}=\varepsilon_{ijk}L_k.
 \label{Coulalg}\ee
Now, let us  perform a simple canonical transformation,
\be
({\bf p}, {\bf r})\to (- {\bf r} , {\bf p}).
 \label{ct}\ee
As a result, the first equation in  \eqref{Coulsur} reads
 \be
  {r}^2-\frac{2{\gamma }}{  p }-2E=0\quad\Rightarrow\quad   p  -\frac{2\gamma}{{r}^2-2E}=0
 \ee
 Interpreting the second equation as an optical Hamiltonian, we get the   refraction index profile known as the ``Maxwell fish eye" \eqref{Mfe0}  with the parameters $\kappa$ and $n_0$ defined as follows
  \be
 \kappa:= -\frac{1}{2E} ,\qquad \frac{n_0}{\lambdabar_0}:=  2\epsilon\kappa\gamma,
 \label{gammaE}\ee
 where $\epsilon=-{\rm sgn}(r^2+1/\kappa)$.

The integrals of motion \eqref{A} result in the symmetry generators of the optical Hamiltonian with the Maxwell fish eye refraction index
\be
\mathbf{L}\to \mathbf{L},\quad \mathbf{A}\to \frac{\mathbf{T}}{2\kappa},\quad \mathbf{T}=\left(1- \kappa  r^2\right)\mathbf{p} + 2 \kappa  (\mathbf{rp}) \mathbf{r}=\left(2-\frac{n_0}{n_{Mfe}({\bf r})}\right)\mathbf{p}+ 2 \kappa  (\mathbf{rp}) \mathbf{r}.
\ee
These integrals form  the $so(4)$ algebra for $\kappa >0$, and $so(1.3)$ algebra for $\kappa<0$:
\be
\{L_i,L_j\}=\varepsilon_{ijk}L_k,\quad \{T_i,L_j\}=\varepsilon_{ijk}T_k,\quad \{T_i,T_j\}=4\kappa \varepsilon_{ijk}L_k\; .
\label{Mfealg}\ee

In the  next sections  we will use the above  described duality for the  construction of the Maxwell fish eye profile for polarized light.

\section{Inclusion of polarization}

Let us briefly discuss the inclusion of polarization.

To  this end we should add to the  scalar Lagrangian $L_0={\bf p\dot{r}}-p+\lambda_0^{-1}n$ the additional term $L_1=-s\mathbf{A}( \mathbf{p})\dot{\mathbf{p}}$,
where $s$ is the spin of the photon, and $\mathbf{ A}$ is the
the vector-potential  of the ``Berry monopole"  (i.e. the
  potential of the magnetic (Dirac) monopole located at the origin of momentum space) \cite{bliokh}
  \be
\mathbf{F}:=\frac{\partial}{\partial \mathbf{ p}}\times \mathbf{A}(\mathbf{p})=\frac{\mathbf{p}}{  {p} ^3}
\label{F}\ee

From the  Hamiltonian viewpoint this means to preserve the form of the Hamiltonian \eqref{h0}
and  replace the  canonical Poisson brackets \eqref{pb0}
by the twisted  ones
 \begin{equation}
  \{x_i, p_j\}=\delta_{ij},\qquad \{x_i, x_j\}=s\varepsilon_{ijk}F_k(\mathbf{p}), \qquad\{p_i, p_j\} =0,
 \label{pbB}\end{equation}
  where $ i,j,k=1,2,3$, and  $F_k$ are  the components of the Berry monopole \eqref{F}.
On this phase space the rotation generators take  the form
  \be
\mathbf{J}=\mathbf{r}\times \mathbf{p}+ s\frac{{\mathbf p}}{ {p} }
\label{Js}
\ee
while
the equations of motion read
\be
\frac{d\mathbf{p}}{dl}=\lambdabar_0^{-1}\mathbf{\nabla} n(\mathbf{r}),\qquad \frac{d\mathbf{r}}{dl}=\frac{\mathbf{p}}{ p } - \frac{s}{\lambdabar_0}\mathbf{F} \times \mathbf{\nabla}n({\bf r}),
  \ee
However, the above procedure, i.e. twisting the Poisson bracket with preservation of the Hamiltonian,  violates  the non-kinematical (hidden) symmetry of the system.
To get the profiles admitting the symmetries in the presence of polarization, we use the following observation \cite{lnp} (see \cite{mardoyan} for its quantum counterpart). Assume we have the three-dimensional rotationally-invariant system
 \be
 \mathcal{H}_0=\frac{{p}^2}{2g(r)}+V(r),\qquad  \{x_i, p_j\}=\delta_{ij}, \quad  \{p_i, p_j\}=\{x_i, x_j\}=0.
 \ee
For the inclusion of  interaction with magnetic monopole, we should  transit from the canonical Poisson brackets to the twisted ones:
\be
\{x_i, p_j\}=\delta_{ij}, \qquad  \{p_i, p_j\}=s\varepsilon_{ijk}\frac{x_k}{r^3},\qquad \{x_i, x_j\}=0.
\label{PBsp}\ee
The rotation generators  then read
\be
\mathbf{J}=\mathbf{r}\times \mathbf{p}+ s\frac{{\mathbf r}}{r}\;: \quad \{J_i,J_j\}=\varepsilon_{ijk}J_k.
\label{Jr}\ee
By modifying the initial Hamiltonian to
\be
\mathcal{H}_s=\frac{{p}^2}{2g( r)}+\frac{s^2}{2g(r){r}^2}+V(r),
\label{Hs}\ee
we find that trajectories of the system preserve their form, but the plane which they belong to, fails to be orthogonal to the
the axis $  {\mathbf{J}} $. Instead, it  turns to the constant angle
\be
\cos\theta_0=\frac{s}{|\mathbf J|}.
\ee
For the systems with hidden symmetries, one can find the appropriate modifications of the hidden symmetry generators respecting the inclusion of the monopole field.

To apply this observation on the systems with polarized light,   we should choose the appropriate integrable system with magnetic monopole, and then perform the canonical transformation
\eqref{ct}
which yields the Poisson brackets for polarized light \eqref{PBsp}. Afterwards we need to solve the following equation
 \be
  r^2+\frac{s^2}{p^2}-2 g(p) (E-V(p))=0,\quad\Rightarrow\quad p=\frac{n^s_{inv} ( r)}{\lambdabar_0}.
 \ee
For example, to get the ``polarized Coulomb profile"  we have  to start  from the free-particle Hamiltonian on three-dimensional  sphere/hiperboloid interacting with Dirac monopole: 
\be
H_{s}= \frac{(1+\kappa r^2)^2}{2}\left({p}^2 +\frac{s^2}{r^2}\right).
\ee
Then, after fixing the energy surface $H_{s}=E$ and performing canonical transformation \eqref{ct}
we arrive to the third-order (with respect to $ {p}^2$) algebraic equation:
   \be
(1+\kappa p^2)^2\left(r^2+\frac{s^2}{p^2}\right)=2E (>0),\qquad{\Leftrightarrow}\qquad  y^3u-y^2(u-\kappa s^2)-Ey+E=0,
\label{nCoul1}\ee
with $ y:=1+\kappa p^2$, $u:=r^2$.

This equation has either one real and two complex solutions or three real solutions, which  describe the ``polarized Coulomb profiles".

Conversely, when we start from the Coulomb problem with Dirac monopole we will arrive to the ``polarized Maxwell fish eye", i.e. the deformation of the ``Maxwell fish eye"  which preserves,  in the presence of polarized light, all symmetries of initial scalar system.
The latter is considered in detail in the next section.

\section{Polarized Maxwell fish eye}
Let us consider  the Coulomb system   with Dirac monopole which is known as ``MICZ-Kepler system" \cite{MICZ}.
It is defined by the twisted Poisson brackets \eqref{PBsp} and  by the Hamiltonian
\begin{equation}
H_{MICZ} =\frac{{  p}^2}{2}+\frac{s^2}{2r^2}-\frac{\gamma}{r}.
\end{equation}
Besides the conserved angular momentum  \eqref{Jr},
this system has the conserved Runge-Lenz vector
\begin{equation}
\mathbf{A}_s=\mathbf{J}\times \mathbf{p}+\gamma \frac{\mathbf{r}}{r},
\label{rls}\end{equation}
which forms the symmetry algebra of Coulomb problem \eqref{Mfealg} (with the replacement $(\mathbf{L},\mathbf{A})\to (\mathbf{J},\mathbf{A}_s)$).
After performing  canonical transformation \eqref{ct}, we get
\be
H_{MICZ}=E\quad\Leftrightarrow \quad r^2+\frac{s^2}{p^2}-\frac{2\gamma}{p}-2E=0.
\ee
Solving this quadratic equation for $p$,  we get the refraction index
given by the expression \eqref{spinFishEye0},
where the notation \eqref{gammaE} is used.
%

The rotation generator \eqref{Jr} transforms to \eqref{Js}, and  the Runge-Lenz vector \eqref{rls} transforms to ${\mathbf{T}_s}/{\kappa}$, where

\be
\mathbf{T}_s= \Big(2-\frac{n_0}{n^s_{Mfe}(\mathbf{r})}\Big)
\mathbf{p}+2\kappa (\mathbf{rp})\mathbf{r} +\frac{2\kappa s}{n^s_{Mfe}(\mathbf{r})} \mathbf{J} .
\ee
Along with \eqref{Js}, these generators form the symmetry algebra of the original Maxwell fish eye profile \eqref{Mfealg} (where the pair $(\mathbf{L},\mathbf{T})$ is replaced by  $(\mathbf{J},\mathbf{T}_s)$).
The Casimirs of the symmetry algebra are given by the expressions
\be
\mathbf{T}^2_s +4\kappa(\mathbf{J}^2-s^2)=\frac{n^2_0}{\lambdabar^2_0}, \qquad \mathbf{T}_s\cdot\mathbf{J}=\frac{s n_0}{\lambdabar_0}.
\ee
Hence, for $\kappa>0$ the vectors  $\sqrt{4\kappa}\mathbf{J}$ and $\mathbf{T}_s$ form the parallelogram with the fixed lengths of  diagonals
\be
 |\mathbf{T}_s\pm \sqrt{4\kappa}\mathbf{J}|=|\frac{n_0}{\lambdabar_0}\pm \sqrt{4\kappa}s|.
\ee
This immediately leads to the conclusion that for $\kappa >0$ the   generators $\mathbf{T}_s $ and $\mathbf{J}$ reach the lower/upper bounds being parallel to each other\be
 \left( |\mathbf{J} |_{\rm min}=s,\; |\mathbf{T}_s|_{\rm max}= \frac{n_0}{\lambdabar_0}\right),\qquad \left(  |\mathbf{J} |_{\rm max}=\frac{n_0}{\lambdabar_0\sqrt{4\kappa}}\;, |\mathbf{T}_s|_{\rm min}= \sqrt{4\kappa}s \right) .
\ee
Notice also, that for   $\kappa > 0$ we get  a  restriction of rays in the finite domain
\be
\kappa > 0\;:\quad r\leq \sqrt{\frac{n^2_0}{4s^2\lambdabar^2_0\kappa^2 }-\frac1\kappa}.
\ee
One can also  note that spin appears in the expression for  the refraction index \eqref{spinFishEye0} along with the factor $\kappa {\lambda_0}^2 = ({\lambda_0}/{2r_0})^2$. In order to stay within the bounds of geometrical optics approximation, this factor must be reasonably small.
Therefore, the influence of the spin will be far more notable within certain range of distance from the core of the fish eye. The latter happens when the condition ${4\kappa s^2\lambdabar^2_0 }/{n_0} \approx n_{Mfe}(\mathbf{r})$ holds.
At these distances the refraction index in presence of spin can be much  smaller as compared to the refraction index with zero spin.
\begin{figure}[hbt!]
\centering
\includegraphics[scale=1.15]{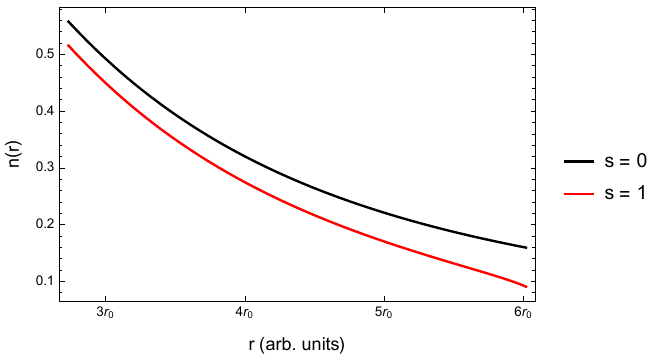}
\caption{Maxwell fish eye refraction index profile for $s=0$ and $s=1$ when $n_0 = 1.5$, $\lambda_0 = 1$, $r_0 = 5$.}
\end{figure}

%

\section{Trajectories}
Let us study the ray trajectories of the polarized light propagating in the medium with above constructed profile \eqref{spinFishEye0}.
One can see that
\be
\mathbf{r}\cdot\mathbf{J}=s\frac{\mathbf{rp}}{p},\quad \mathbf{r}\cdot\mathbf{T}_s = \frac{n_0}{\lambdabar_0}\frac{\mathbf{rp}}{p}\qquad\Rightarrow \qquad \mathbf{r}\cdot\left(\mathbf{J}-
\frac{s\lambdabar_0}{n_0}\mathbf{T}_s\right)=0.
\ee
Hence, ray trajectories are orthogonal  to the axis
\be
\mathbf{E}_3=\mathbf{J}-\frac{s\lambdabar_0}{n_0}\mathbf{T}_s,
\ee
and, therefore the trajectories belong  to the plane spanned by the following vectors
\be
\mathbf{E}_1=\mathbf{T}_s\times\mathbf{J},\qquad \mathbf{E}_2=\mathbf{E}_3\times\mathbf{E}_1=\Big(\mathbf{J^2}-s^2\Big)\left(\mathbf{T}_s -  \frac{4 s\lambdabar_0\kappa}{n_0}\mathbf{J}\right)
 \quad :\quad \mathbf{E}_3\cdot\mathbf{E}_2=\mathbf{E}_3\cdot\mathbf{E}_1=0.
\label{eee}
\ee
Then, from the expression $\mathbf{J}\cdot\big(\mathbf{r}\times\mathbf{T_s}\big)$ we  immediately obtain the solution  for the ray trajectories:
\be
\mathbf{r}\cdot\big(\mathbf{T_s}\times\mathbf{J}\big)=\Big(J^2-s^2\Big) \Big(2 - \frac{n_0}{n^{s}_{mfe}}\Big).
\label{traj}
\ee
This prompts us to  introduce the following orthogonal frame
\be
\mathbf{e}_i=\frac{\mathbf{E}_i}{|\mathbf{E}_i|}\; :\quad  \mathbf{e}_i\cdot\mathbf{e}_j=\delta_{ij},
\ee
where
\be
|\mathbf{E}_1|^2=\left(\mathbf{J}^2-s^2\right)\left(\frac{n^2_0}{\lambdabar^2_0}-{4\kappa \mathbf{J}^2}\right),\qquad |\mathbf{E}_3|^2=\left(\mathbf{J}^2-s^2\right)\left(1-\frac{4 s^2\lambdabar^2_0\kappa}{n^2_0}\right),\qquad |\mathbf{E}_2|^2=|\mathbf{E}_1|^2|\mathbf{E}_3|^2
\label{emod}\ee
Decomposing $\mathbf{r}$ over this frame, we introduce the  polar coordinates
\be
\mathbf{r}=x_1\mathbf{e}_1+x_2\mathbf{e}_2,\qquad x_1=r\cos\varphi,\quad x_2=r\sin\varphi.
\ee
Then,
having in mind Eqs. \eqref{eee} and \eqref{emod}, we can immediately rewrite  the  equation \eqref{traj} in polar coordinates
\be
1-  |\kappa||\mathbf{a}_s|r\cos\varphi=\frac{1+\kappa r^2}{1+\sqrt{1-\frac{4\kappa s^2\lambdabar^2_0}{n^2_0}(1+\kappa r^2)}},
\ee
where
\be
|\mathbf{a}_s|^2:=R^2_s-\frac 1\kappa,\qquad  R^2_s:=\frac{n^2_0-4\kappa s^2\lambdabar^2_0}{4\lambdabar^2_0\kappa^2(J^2-s^2)}.
\label{attr}\ee
So the trajectories of polarized light are not circles anymore, in contrast to the case of scalar waves. However, they can be attributed by the parameters ${\bf a_s}$ and $R_s$  \eqref{attr} which, in the limit $s\to 0$, become the center coordinate and the radius of the circle, respectively. Indeed, for $s=0$  the solution \eqref{traj} results in the   the equation for circle with the center located at $\mathbf{e}_1$ axis
\be
s=0\; :\quad \mathbf{r}\cdot\big(\mathbf{T}\times\mathbf{L}\big)=L^2(1-\kappa r^2) \quad\Rightarrow\quad \left(\mathbf{r}-\mathbf{a}_0\right)^2=R^2_0,
\ee
where
  \be \mathbf{a}_0:=\frac{\mathbf{T}\times\mathbf{L}}{2\kappa L^2},\quad |\mathbf{a}_0|^2=R^2_0-\frac{1}{\kappa},\quad R_0:=\frac{n_0}{2|\kappa|\lambdabar_0 L}.
\ee
\begin{figure}[hbt!]
\centering
\includegraphics{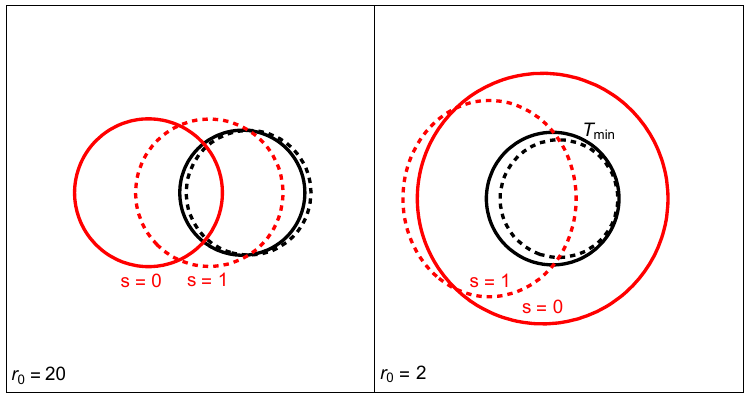}
\caption{Deformations of the ray trajectories for different values of $r_0$ when $n_0 = 1.5$, $\lambdabar_0 = 1$. The black curves correspond to the basic trajectories where $T = T_{min} = s/r_0$. The red (light gray) curves correspond to trajectories with intermediary value of $T$. Dashed curves are the trajectories corresponding to the same value of $T$ but for circularly polarized light ($s = 1$). The first figure ($r_0 = 20$) corresponds to the case when the deformations of the profile only result to the shift of the centers of the trajectories not affecting their shapes. Conversely, in the second figure ($r_0 = 2$), the deformations of the Maxwell fish eye result to highly deformed trajectories.}
\end{figure}
Given $s\lambdabar_0\ \ll n_0/\kappa$, the deformation of circle is negligible, which is not the case for $s\lambdabar_0\  \sim n_0/\sqrt{\kappa}$.

As we can see from Fig.2, for $r_0 = 20$ the only notable manifestation of the polarization is the shift of the center of the trajectory.
However, it is worth noting that since the dashed lines are not circles anymore when talking about the center of the trajectory for  $s = 1$, we refer to the point which becomes the center of the circular trajectory when we pass from $s = 1$ to $s = 0$.
The second picture illustrates the circular trajectories and their deformations for  $s\lambdabar_0\  \sim n_0/\sqrt{\kappa}$. In this case,  the original  profile \eqref{Mfe0} and the deformed one  \eqref{spinFishEye0} differ drastically. The circular trajectories are notably deformed.

{  Detailed knowledge of trajectory parameters \eqref{attr} can be used in different applications. For example, in the conform mapping scheme the cloaking area is the outer space of closed trajectories \cite{Leonhardt06}. Therefore as it follows from \eqref{attr} there is no cloaking for polarized photons when $J\to s$.}

\section{Concluding remarks}
The standard Maxwell fisheye profile does not ensure closed ray trajectories for polarized photons (the only exception are linearly polarized photons corresponding to the $s=0$ spin stat), while the closeness of trajectories is the main property that is used in perfect imaging and cloaking phenomena.
In this paper, we suggested the deformation of the Maxwell fisheye profile which ensures the closeness of the trajectories of the ray trajectories for the polarized photons. We examined the properties of the deformed profile and have shown that the main difference between the cases of polarized and non-polarized photons is observed at the vicinity of wave and geometrical optics border $s\lambda_0/r_0\sim 1$.

Proposed modification scheme is applicable  for any isotropic refraction index $n(r)$. Namely, to preserve the qualitative properties of scalar wave trajectories for the propagating polarized light, we should replace it with the modified index $n^s(r)$ which is the solution
(with respect to $p$) of the following equation:
\be
p=\frac{1}{\lambdabar_0} n\left(\sqrt{r^2+\frac{s^2}{p^2}}\right),\quad\Rightarrow\quad p=n^s(r),
\ee
where $s$ is polarization of light.
The proposed deformation preserves the additional symmetries of the system (if any), and thus, guarantees the closeness of trajectories of polarized light.

Seemingly, the suggested scheme could be extended to some non-isotropic, but integrable profiles as well. On the other hand, non-isotropic integrable profiles are not common objects in the present study, though they obviously can be constructed by the use of existing integrable models.  For example,  choosing a textbook integrable system, the two-center Coulomb problem \cite{arnold} and performing trivial canonical transformation \eqref{ct} we can construct (taking into account the  expressions for constants of motion, see, e.g. \cite{2C}) anisotropic profile which could be interpreted as a superposition of two ``Maxwell fish eye" profiles. Furthermore, using the proposed scheme,  we can construct ``polarized Maxwell double fish-eye" profile as well, starting from the ``two-center MICZ-Kepler problem" \cite{kno}, i.e. from the two-center Coulomb problem specified by the presence of magnetic monopoles located at the attraction  centers.
We hope to consider this problem elsewhere.
%
%
%
%

\acknowledgements
Authors acknowledge partial  financial support from
Armenian Committee of Science, projects 20RF-023 (A.N., Zh.G.), 21AG-1C062  and  from the Russian Foundation
of Basic Research   grant 20-52-12003 (A. N.). This work was done within
 ICTP projects   NT-04 and   AF-04.

\end{document}